# Weak Interaction and Cosmology

P. R. Silva -  Departamento de Física – ICEx – UFMG
CP – 702 - 30123-970 – Belo Horizonte – MG – Brazil
e-mail: prsilvafis@terra.com.br

ABSTRACT- In this letter we examine the connection among the themes: the cosmological constant, the weak interaction and the neutrino mass. Our main propose is to review and modify the ideas first proposed by Hayakawa [ Prog. Theor. Phys.**Suppl.**,532(1965).], in the ligth of the new-fashioned features of contemporary physics. Assuming the pressure of a Fermi gas of neutrinos should be balanced by its gravitational attraction, we evaluate the mass of the background neutrino and its number.The neutrino mass here evaluated is compatible with the known value for the cosmological constant (or dark energy).Taking in account the role played by the weak forces experimented by the neutrinos, we also determined a value for the electroweak mixing angle. For sake of comparison, an alternative evaluation of the neutrino mass is also done.

The cosmological constant problem [1] has challenged people working in theoretical physics through the years. The recent discovery of the dark energy [2,3] and its possible conection  with the cosmological constant and the neutrino mass has increased the interest in the study of these subjects [4,5,6].

On the other hand, as was pointed out by Okun [7], neutrino masses manifest themselves in experiments devoted to study neutrino oscillations (se for example [8]), whose searchs were first proposed by Pontecorvo [9], see also [10].

Fordon, Nelson and Weiner [11] advanced the idea that the dark energy sector is tied to that of neutrinos. This purpose was further elaborated by Peccei [12]. Sidharth [13] claims that the neutrino mass, roughly a billionth that of the electron, was theoretically deduced by him [14]. In his paper entitled: " Cosmological Interpretation of the Weak Forces ", Satio Hayakawa [15] postulates that: the weak forces are dictated by cold neutrinos of finite mass which fills the universe ( see also: " Atomism and Cosmology ", by the same author [16] ).

In this letter, we intend to examine the connection among these three themes, namely: the cosmological constant, the weak interaction and the neutrino mass. Our main propose is to review and to modify the ideas first introduced by Hayakawa [15,16], in the ligth of  new-fashioned features of  contemporary physics.

As a starting point, let us review the deduction of Hsu and Zee [4] of the cosmological constant parameter. As pointed out by Hsu and Zee [4], one of the aspects of the cosmological constant puzzle is that, -$\Lambda$ couples to the space-time volume $\int d^4x \sqrt{-g}$ of the universe. This contributes to an effective action $S_{eff}$ with a leading term $-\Lambda R^4$, where R is the radius of the observable universe. This term behaves smoothly when $\Lambda$ goes through zero, from positive to negative values. Also, according to  Hsu and Zee [4], we need a term which displays a singular behavior, when $\Lambda$ goes through zero  and suggests that quantum gravity produces a term of the form $- M_{Pl}^4 / \Lambda$, resulting in the effective action

$$S_{eff} \sim - ( \Lambda R^4 + M_{Pl}^4/\Lambda ) + \Lambda( \text{independent terms} ). \quad (1)$$

Extremizing this action with respect to $\Lambda$ leads to

$$\Lambda \sim ( M_{Pl} /R )^2 \sim ( L_{Pl} R )^{-2}. \quad (2)$$

The above result also can be obtained through the holographic dark energy model [17,18,19]. A. Cohen et al. [17] suggested earlier that the zero-point energy in quantum field theory is affected by an infrared cutt-off L. If L is identified with $R = H^{-1}$, where H is the Hubble constant, they obtain relation (2) for the cosmological constant. Relation (2) is only an order of magnitude estimate of $\Lambda$, neglecting numerical factors that we suppose to be of order one. In the following we are going to make an evaluation of the neutrino mass, starting from cosmological arguments. As we will see, the value that we will obtain for this quantity, is closely related to the cosmological constant $\Lambda$ found in previous works [4,17,19].

One of the hypothesis considered by Hayakawa [16] is that the balance between particles and antiparticles in the universe is not necessarily satisfied by each kind of particles but for all kind of particles. According to Pontecorvo [20], if the universe is filled with neutrinos and anti-neutrinos whose number is far greater than the number of all other particles, the balance between particles and anti-particles holds with sufficient accuracy. We also suppose that these background neutrinos form a degenerate Fermi gas. The pressure of this Fermi gas should be balanced by the gravitational attraction caused by the background neutrinos. However, instead to take the non-relativistic approximation for Fermi energy as in references [15,16], we will consider the Fermi energy $E_F$ in its relativistic form, namely

$$E_F = (\hbar c / R) N_\nu^{1/3} (3/\pi)^{1/3} . \qquad (3)$$

In (3), $\hbar$ is the reduced Planck constant, c the speed of ligth and $N_\nu$ the number of background neutrinos in the universe. The gravitational energy is given by

$$E_G = - ( G N_\nu m_\nu^2 )/R, \qquad (4)$$

where G is the gravitational constant and $m_\nu$ the neutrino mass. We observe that taking in account tthe relativistic character of neutrinos, its speed being very close to c [7], may be that $m_\nu$ is indeed another way of writting $E_\nu/c^2$, where $E_\nu$ is the neutrino energy. Therefore, the effective energy can be written as

$$E_{eff} = (\hbar c / R) N_\nu^{1/3} (3/\pi)^{1/3} - ( G N_\nu m_\nu^2 )/R. \qquad (5)$$

Taking the extremum of (5) with respect to $N_\nu$, we obtain

$$m_\nu^2 = [ \hbar c (3/\pi)^{1/3}]/(3 G N_\nu^{2/3}). \qquad (6)$$

However in (6) $m_\nu$ is a function of $N_\nu$, a quantity yet to be determined. Overcoming this difficulty requires another independent determination of $m_\nu$. In the following we are trying to do this.

In reference [21], an extension of the MIT bag model [22] developed to describe the strong interaction inside the hadronic matter was proposed, as a means to account for the confinement of matter in the universe. This model of the bag-universe is described by the potential $V_{u-bag}$,

$$V_{u-bag} = 2 G M^2/R + (4/3) \pi R^3 P. \qquad (7)$$

In (7), M is the mass of the universe and P is the pressure of vacuum at the boundary of the universe. The first term of (7) accounts for the gravitational attraction, and to justify the second contribution, we borrow a reasoning given by Jaffe [23]: " It is possible to distinguish an otherwise empty region of the space in a way consistent with relativity by subjecting the boundary of the region to a constant pressure P, where P may be thought of as a pressure exerted by the otherwise empty space on the universe.". Here we have adapted Jaffe's [23] reasoning to the bag-universe.

Differentiating $V_{u\text{-bag}}$ with respect to R and making the requirement that its minimum value equals to the "rest" energy of the universe we obtain

$$P = [9/(128\pi)] (c^2 H_0)/G \qquad (8)$$

Taking $H_0$, the Hubble "constant" as $3 \times 10^{-18}$ s$^{-1}$ [24], we get P = $2.7 \times 10^{-15}$ atm, the pressure at the boundary of the universe, which could be compared to $8 \times 10^{28}$ atm at the boundary of the nucleon, as quoted by Jaffe [23].

We may consider that the pressure of vacuum P confines a gas of background neutrinos and we write

$$P V = (1/2) N_\upsilon m_\upsilon c^2, \qquad (9)$$

where $V = (4/3) \pi R^3$ is the volume of the bag-universe. From (9), we can write:

$$m_\upsilon = (8\pi/3) [(P R^3)/( N_\upsilon c^2 )], \qquad (10)$$

where P is given by (8). Comparing (10) and (6) give us:

$$N_\upsilon = (27/256)^{3/4} (\pi/3)^{1/4} (R/L_{Pl})^{3/2} = f (R/L_{Pl})^{3/2}, \qquad (11)$$

where we have introduced the number f as a matter of convenience, and

$$m_\upsilon = [3/(16f)] (\hbar/c) R^{-1/2} L_{Pl}^{-1/2}. \qquad (12)$$

Numerical evaluation of (11) and (12), by taking R = $10^{26}$ m and $L_{Pl}$ = $10^{-35}$ m leads to:

$$N_\upsilon = 5.9 \times 10^{90}, \qquad (13)$$

and

$$m_\upsilon = 6.3 \times 10^{-3} \text{ ev}/c^2. \qquad (14)$$

The values for $N_\upsilon$ and $m_\upsilon$ estimated here are of the same order of magnitude as those estimated by Hayakawa [15,16], namely $10^{91}$ and $10^{-2}$ ev/c$^2$ respectively. The numerical factor $3/(16f)$ is close to 1, and therefore the background neutrino mass evaluated in this work is of the same order of magnitude as $M_\Lambda$ from the work of Hsu and Zee [4], which is related to $\Lambda$ by:

$$\Lambda = M_\Lambda^4. \qquad (15)$$

Hayakawa [15,16] has considered that the competition between Fermi repulsion and gravitational attraction leads the neutrino gas to undergo an oscillation with a certain angular frequency ω. The zero-point energy of this oscillation is related to the weak interaction through the equation

$$(1/2)\,\hbar\omega = G_F\,N_\upsilon/V, \qquad (16)$$

where $G_F$ is the Fermi coupling constant.

In the present work it is not possible to evaluate a zero- point energy in the way proposed by Hayakawa. However, Volovik [25] has proposed a connection between the problems of the vacuum energy of quantum hydrodynamics and the cosmological constant.

On the other hand, it is well known that both in superconductors and superfluid systems, their wave functions can extend its coherence over macroscopic distances (scales). Therefore we will adopt the hypothesis that the range of the weak interactions can be extended to the scale of length of the cosmological constant and we write:

$$\hbar\omega = \alpha_w(m_\upsilon)\,\hbar c/l_\upsilon, \qquad (17)$$

where

$$\alpha_w(m_\upsilon) = \alpha\,(m_\upsilon/M_w)^2, \qquad (18)$$

and

$$l_\upsilon = \hbar/(m_\upsilon c). \qquad (19)$$

In (17), $\alpha_w(m_\upsilon)$ is the weak coupling constant evaluated at the mass scale $m_\upsilon$, and $l_\upsilon$ is the length scale associated to this mass (its Compton length), and $M_w$ is the mass of the charged boson which intermediates the weak interaction, the so-called intermediate vector boson.

From (16) we can write

$$G_F = (1/2)\,\hbar\omega\,V/N_\upsilon, \qquad (20)$$

and by considering (11), (12), (17), (18) and (19), we finally obtain

$$G_F/(\hbar c)^3 = (2^{13}/3^6)\,\alpha\,(M_w c^2)^{-2}. \qquad (21)$$

This result must be compared with [26]

$$G_F/(\hbar c)^3 = (\pi/\sqrt{2})\,\alpha\,(\sin^2\theta_w)^{-1}\,(M_w c^2)^{-2}. \qquad (22)$$

Indeed, relation (22) displays the dependence of the Fermi coupling $G_F$ on $\theta_w$, the electroweak mixing angle. According to Kane (see chapters 6 and 7 of reference [10]), $\theta_w$ must be measured or calculated in some way independent of the electroweak theory of Glashow [27], Salam [28] and Weinberg [29]. Comparing (21) and (22), yields

$$\sin^2\theta_w = (3^6/2^{13})\,(\pi/\sqrt{2}) \approx 0.198. \tag{23}$$

This estimate of $\sin^2\theta_w$ must be compared with the values

$$\sin^2\theta_w = 0.230 \pm 0.005 \text{ ( from reference [30] ) , and}$$

$$\sin^2\theta_w = 0.219 \pm 0.007 \pm 0.018 \text{ ( from reference [31] ),}$$

the last one refering to a precision measurement of parity non-conservation in atomic Cesium, see also [32] and [33]. We observe that the results of reference [30] comes from high-energy measurements, while in reference [31], the measurement was performed at a fraction of electronvolts.

Until now we have tied the neutrino mass to the energy scale of the cosmological constant, and these considerations lead to the possibility of linking weak interactions and cosmology. On the other hand, Zee [34] in his recent book on field theory, proposes a line of attack as a means to construct an effective field theory of a neutrino mass. As was pointed out by Zee [34], the way of doing this is to construct an SU(2)×U(1) invariant effective theory. In order to do this, a dimension-5 operator is required. Yet according to Zee: schematically a $l_L l_L$ contribution to the Lagrangian contains the desired neutrino bilinear ( L refers to left ) but carries hypercharge Y/2 = -1; on the other hand, the Higgs doublet φ carries hypercharge + 1/2, and so the lowest dimensional operator that we can form is like llφφ with dimension 5. Thus, the effective Lagrangian must contain a term (1/M)llφφ, with M the mass scale of the new physics responsible for the neutrino mass.

In the following we propose a tentative estimation of the order of magnitude of this mass scale. In the Dirac's extensible model of the electron [35] (see also [36] ), it is possible to associate a radius related to its self electromagnetic interaction, the so-called classic radius of the electron. Alternatively, we could think in a weak radius of the electron, and we write

$$m_e c^2 = \alpha_w(m_e)\,\hbar c / (2R_{ew}), \tag{24}$$

where

$$\alpha_w(m_e) = \alpha\,(m_e/M_w)^2. \tag{25}$$

Numerical evaluation of (24) gives for the weak radius of the electron the value

$$R_{ew} = 5 \times 10^{-26}\,\text{m}. \tag{26}$$

Associated to this radius, we have a mass scale of

$$Mc^2 = \hbar c / (R_{ew}) = 4 \times 10^9\,\text{Gev}. \tag{27}$$

Now, let us turn again to the Zee [34] reasoning. By dimensional analysis, we can estimate

$$m_\upsilon \sim m_l^2 / M. \tag{28}$$

If we take $m_l \sim 10^2$ Mev to be the muon mass and the value of M evaluated in (27), we get

$$m_\upsilon \sim 10^{-3} \text{ ev}/c^2. \qquad (29)$$

Incidentally a weak radius also can be attributed to the muon, of the order of magniitude of $10^{-24}$ m, and the mass scale associated to this radius is $\sim 10^7$ Gev.

It is worth to point out that, in a somewhat different way, Zee [34] postulated a value of $10^{-1}$ ev for the mass energy of neutrino, and by using (28), he found M $\sim 10^8$ Gev as the energy scale of the "new physics".

ACKNOWLEDGEMENTS – We are grateful to Domingos Sávio de Lima Soares, Antônio Paulo Baêta Scarpelli e Marcos Sampaio for reading the manuscript.